\documentclass[aps,prl,superscriptaddress,twocolumn]{revtex4}

\usepackage{bm}
\usepackage{graphicx}

\begin{document}

\title{Dissipation-driven phase transition in two-dimensional
       Josephson arrays}

\author{Luca Capriotti}
\affiliation{Valuation Risk Group, Credit Suisse First Boston (Europe) Ltd.,
             One Cabot Square, London E14 4QJ, United Kingdom}
\affiliation{Kavli Institute for Theoretical Physics, University
    of California, Santa Barbara, CA 93106, USA}
\author{Alessandro Cuccoli}
\affiliation{Dipartimento di Fisica dell'Universit\`a di Firenze -
             via G. Sansone 1, I-50019 Sesto Fiorentino (FI), Italy}
\affiliation{Istituto Nazionale per la Fisica della Materia
             \,-\,U.d.R. Firenze -
             via G. Sansone 1, I-50019 Sesto Fiorentino (FI), Italy}
\author{Andrea Fubini}
\affiliation{Dipartimento di Fisica dell'Universit\`a di Firenze -
             via G. Sansone 1, I-50019 Sesto Fiorentino (FI), Italy}
\affiliation{Istituto Nazionale per la Fisica della Materia
             \,-\,U.d.R. Firenze -
             via G. Sansone 1, I-50019 Sesto Fiorentino (FI), Italy}
\author{Valerio Tognetti}
\affiliation{Dipartimento di Fisica dell'Universit\`a di Firenze -
             via G. Sansone 1, I-50019 Sesto Fiorentino (FI), Italy}
\affiliation{Istituto Nazionale per la Fisica della Materia
             \,-\,U.d.R. Firenze -
             via G. Sansone 1, I-50019 Sesto Fiorentino (FI), Italy}
\affiliation{Istituto Nazionale di Fisica Nucleare,
             Sezione di Firenze}
\author{Ruggero Vaia}
\affiliation{Istituto dei Sistemi Complessi
             del Consiglio Nazionale delle Ricerche, Sezione di Firenze,
             via Madonna del Piano, I-50019 Sesto Fiorentino (FI), Italy}
\affiliation{Istituto Nazionale per la Fisica della Materia
             \,-\,U.d.R. Firenze -
             via G. Sansone 1, I-50019 Sesto Fiorentino (FI), Italy}
\date{\today}

\begin{abstract}
We analyze the interplay of dissipative and quantum effects in the
proximity of a quantum phase transition. The prototypical system is a
resistively shunted two-dimensional Josephson junction array, studied
by means of an advanced Fourier path-integral Monte Carlo algorithm.
The reentrant superconducting-to-normal phase transition driven by
quantum fluctuations, recently discovered in the limit of infinite
shunt resistance, persists for moderate dissipation strength but
disappears in the limit of small resistance.  For large quantum
coupling our numerical results show that, beyond a critical
dissipation strength, the superconducting phase is always stabilized
at sufficiently low temperature.  Our phase diagram explains recent
experimental findings.
\end{abstract}


\maketitle

Dissipation due to the coupling with the surrounding
environment~\cite{CaldeiraL1983} is an unavoidable effect
accompanying the operation of any microscopic or mesoscopic quantum
device. The knowledge of the influence of dissipative effects on
quantum coherence and quantum phase transitions (QPT) is therefore
essential to assess the reliability of such devices in performing
tasks which strongly depend on the possibility to maintain
entanglement (phase coherence), e.g., in quantum computation. Among
the quantum devices that have already found wide application, many
are based on a collection of regularly arranged or single small
Josephson junctions~\cite{SchonZ1990,FazioZ2001}. These are also
among the candidates for the physical implementation of the {\em
so-called} q-bits~\cite{MakhlinSS2001}.

Josephson junction arrays (JJA), are prototypical systems displaying
a quantum phase transition with a control parameter tuning the
strength of quantum fluctuations. This has become progressively clear
after it was pointed out in the late 70's that the charging energy of
Josephson-coupled superconducting grains could lead to the quenching
of the collective superconducting phase. Since then, several
studies~\cite{SchonZ1990,FazioZ2001} have been devoted to
characterizing the superconductor-normal (SN) transition in JJA.
Among the systems studied, the two-dimensional (2D) ones are the most
interesting as no true long-range order is possible at finite
temperature while a genuine QPT occurs at
$T\,{=}\,0$~\cite{SondhiGCS97}; moreover, 2D samples can be
fabricated in a controlled way and experimentally
characterized~\cite{ZantEGM1996,RimbergHKCCG1997,TakahideYKOK2000}.

The main effect of dissipation in JJA is that of quenching the
quantum fluctuations of the phase variables (thereby enhancing those
of the conjugate charges) thus stabilizing the S
phase~\cite{SimanekB1986,ChakravartyIKL1986,Fisher1987,
ChakravartyIKZ1988,ChoiJ1989,KimC1995,PanyukovZ1989}. This has had
ingenious experimental confirmations, e.g., in a JJA coupled with a
2D electron gas substrate that allows one to tune the dissipation
strength~\cite{RimbergHKCCG1997}, as well as in identical JJA with
different built-in Cr shunt resistors~\cite{TakahideYKOK2000}.

In this paper, we use a Fourier path-integral Monte Carlo (PIMC)
approach to analyze the competition between dissipation and quantum
fluctuations in JJA, for strong quantum coupling, and in proximity of
the QPT. In particular, we study the phase diagram as a function of
temperature, quantum coupling and dissipation. We show how a large
enough dissipation leads to the disappearance of the zero-temperature
QPT, as well as of the reentrant low-temperature behavior displayed
in the limit of large shunt resistance~\cite{CCFTV2003}. The phase
diagram we obtain is in remarkable agreement with the experimental
one. Concerning the reentrance, we can explain the two distinct
behaviors found in experiments; indeed, a non-monotonic (i.e.,
reentrant) low-temperature behavior of the array resistance has been
observed in unshunted JJA~\cite{ZantEGM1996}, but not in shunted
ones~\cite{TakahideYKOK2000}.

JJA are essentially described by the quantum $XY$ model, whose
coordinates and momenta correspond to the wave-function phases
$\hat\varphi_{\bm{i}}$ and the net Cooper-pair number (charge)
$\hat{n}_{\bm{i}}$, respectively, of superconducting islands arranged
on a lattice:
\begin{equation}
 \hat{\cal{H}} = \frac{(2e)^2}2\sum_{\bm{ij}}
 C_{\bm{ij}}^{-1}\,\hat{n}_{\bm{i}}\hat{n}_{\bm{j}}
 - \frac{E_{_{\rm{J}}}}2 \sum_{\bm{id}}
 \cos (\hat\varphi_{\bm{i}}-\hat\varphi_{\bm{i+d}}) ~,
\label{e.H}
\end{equation}
where
$[\hat\varphi_{\bm{i}},\hat{n}_{\bm{j}}]\,{=}\,i\,\delta_{\bm{ij}}$
and $\bm{d}$ runs over the $z$ nearest-neighbor displacements. The
Josephson energy $E_{_{\rm{J}}}$ sets the energy scale, making it
convenient to use the dimensionless temperature
$t\,{=}\,T/E_{_{\rm{J}}}$, while the charging energy involves the
capacitance matrix $C_{\bm{ij}}=C\,\Gamma_{\bm{ij}}^{(\eta)}$, with
$\Gamma_{\bm{ij}}^{(\eta)}=\big(z\,\delta_{\bm{ij}}
-{\textstyle\sum_{\bm{d}}}\delta_{\bm{i},\bm{j+d}}\big)
+\eta\,\delta_{\bm{ij}}$ including the mutual capacitance $C$ and the
self capacitance $C_0\,{\equiv}\,\eta\,C$; in the experimental
samples $C$ is typically
dominant~\cite{ZantEGM1996,TakahideYKOK2000}, i.e., $\eta\,{\ll}\,1$.
We consider here a square-lattice, $z\,{=}\,4$.

In the classical limit (large $C$) the charging term is
thermodynamically irrelevant and the SN transition is a
Berezinskii-Kosterlitz-Thouless (BKT) one~\cite{BKT}, leading from a
high-temperature disordered phase to a low-temperature quasi-ordered
phase at $t_{_{\rm{C}}}\,{=}\,0.892$~\cite{Olsson1994}. The
superconducting phase is weakened~\cite{FazioZ2001} by the quantum
fluctuations of the phase when the characteristic charging energy
$E_{_{\rm{C}}}\,{=}\,(2e)^2/2C$ is comparable to the Josephson energy
$E_{_{\rm{J}}}$, i.e., when the quantum coupling constant
$g\,{=}\,\sqrt{E_{_{\rm{C}}}/E_{_{\rm{J}}}}$ is of the order of
unity~\cite{CFTV2000}. The coupling parameter $g$ can be varied in
the fabrication of JJA, and its further increase can finally drive
the zero-$T$ system through a QPT at a critical value
$g_{_{\rm{C}}}$, numerically~\cite{RojasJ1996-BelousovL1996}
estimated ${\simeq}\,3.4$. Near the QPT the system is characterized
by a sharp enhancement of anharmonic quantum
fluctuations~\cite{CCFTV2003}.

\begin{figure}[t]
\includegraphics[width=85mm]{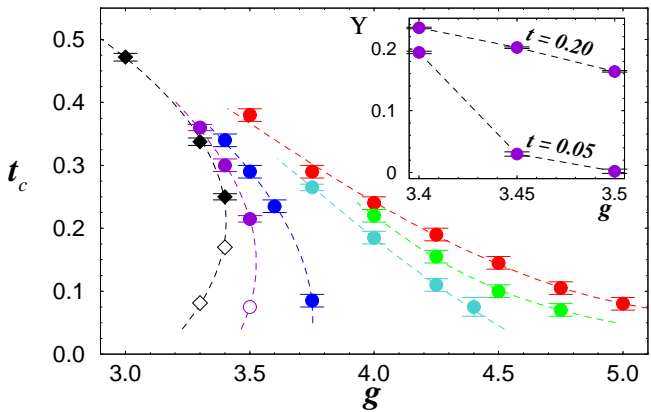}
\caption{
Phase diagram for the square-lattice JJA with $\eta\,{=}\,0.01$ for
increasing values of the dissipation strength
$\gamma\,{=}\,R_{\rm{Q}}/R_{\rm{S}}\,{=}\,0$ (diamonds), 0.15, 0.25,
0.4, 0.45, and 0.5 (circles, from left to right). Inset: Trotter- and
size-extrapolated helicity modulus $\Upsilon(g,t)$ for
$\gamma\,{=}\,0.15$, at $t\,{=}\,0.2$ and $t\,{=}\,0.05$.
\label{f.1_Tg_gamma}}
\end{figure}

In this paper we consider normal Ohmic shunt resistors $R_{_{\rm S}}$
as the source of dissipation, whose strength is measured by the
dimensionless parameter $\gamma\,{=}\,R_{\rm{Q}}/R_{_{\rm S}}$, with
the resistance quantum $R_{\rm{Q}}\,{=}\,h/(2e)^2$. It is well known
that dissipation enhances the S phase, i.e.,
$t{_{_{\rm{C}}}}(g,\gamma)$ increases with $\gamma$, but reliable
results could be obtained only in the low-coupling
regime~\cite{KimC1995,CFTV2000}. Mean field, renormalization group,
and variational approaches~\cite{SimanekB1986,ChakravartyIKL1986,
Fisher1987,ChakravartyIKZ1988} predict the existence of a critical
value $\gamma{_{_{\rm{C}}}}\,{=}\,2/z\,{=}\,1/2$ above which the QPT
disappears and for any value of $g$ the system is in the S phase at
sufficiently low temperature. Moreover, it is not clear whether the
reentrant behavior (with the N phase reappearing at lower $T$)
observed at $\gamma\,{=}\,0$ in the proximity of
$g_{_{\rm{C}}}$~\cite{CCFTV2003} disappears before or at the critical
value $\gamma{_{_{\rm{C}}}}$. As these phenomena involve the strong
coupling region where approximate theories give contradictory
answers~\cite{theories}, accurate numerical data are required for a
real understanding, as well as for the interpretation of the
experimental data.

To this purpose we employ an efficient Fourier PIMC technique
recently introduced~\cite{CCFTV2002,CCFTV2003}. This starts from the
path-integral for the effective partition function
${\cal{Z}}=\oint{\cal{D}}{\bm\varphi}
\exp\big\{-S[\bm\varphi]-S_{_{\rm{I}}}[\bm\varphi]\big\}$,
where the action $S[\bm\varphi]$ corresponds to the
Hamiltonian~(\ref{e.H}) and $S_{_{\rm{I}}}[\bm\varphi]$ is the
bilocal {\em influence}
action~\cite{CaldeiraL1983,Weiss1999,ChakravartyIKL1986}, namely
\begin{eqnarray}
 S[\bm\varphi]\!= \!\!\!\int\limits_0^{1/t}\!\!
 du \bigg[\!\sum_{\bm{ij}} \frac{\Gamma_{\bm{ij}}^{(\eta)}}{4g^2}\,
 {\dot \varphi_{\bm{i}}(u)\,\dot\varphi_{\bm{j}}(u)}
 {-} \frac12\!\sum_{\bm{id}}
  \cos\varphi_{\bm{id}}(u) \bigg], &&
 \label{e.S}
\\
 S_{_{\rm{I}}}[\bm\varphi]= \frac12\!
 \int\limits_0^{1/t}\! du\,du'~\kappa(u{-}u')
 \sum_{\bm{id}}\big[\varphi_{\bm{id}}(u){-}\varphi_{\bm{id}}(u')
 \big]^2~. &&
\label{e.SI}
\end{eqnarray}
Here $u\in[0,1/t]$ is the (dimensionless) `imaginary time',
$\varphi_{\bm{id}}\,{=}\,\varphi_{\bm{i}}\,{-}\,\varphi_{\bm{i+d}}$,
and $\kappa(u){=}\gamma\,t^2/8\sin^2(\pi tu)$ is the dissipative
Ohmic kernel. Using Fourier-Matsubara variables
it is easily shown that the influence action turns into a local form,
suggesting that dissipation can be easier dealt with in Fourier
space. Therefore, at variance with the standard PIMC algorithm, that
samples the variables
$\big\{\varphi_{\bm{i}\ell}\,{=}\,\varphi_{\bm{i}}(\ell/Pt)~,
~\ell\,{=}\,1,\,...,\,P\big\}$ after discretization of the interval
$u\in[0,1/t]$ in $P$ slices of size $1/Pt$ ($P$ being the Trotter
number), we proposed~\cite{CCFTV2002} to sample the $P$ Fourier
components of $\varphi_{\bm{i}\ell}$. Choosing $P\,{=}\,2M{+}1$ one
can write
\begin{equation}
 \varphi_{\bm{i}\ell} = \bar\varphi_{\bm{i}}
 +2\sum_{k=1}^M \bigg(\varphi_{\bm{i}k}^{^{\rm{(R)}}}
 ~\cos\frac{2\pi\ell k}P
 +\varphi_{\bm{i}k}^{^{\rm{(I)}}}
 ~\sin\frac{2\pi\ell k}P\bigg) ~.
\label{e.Fphi}
\end{equation}
The $2M$ components $\big\{\varphi_{\bm{i}k}^{^{\rm{(R)}}},
\varphi_{\bm{i}k}^{^{\rm{(I)}}}\big\}$ and the
zero-frequency component $\bar\varphi_{\bm{i}}$ are sampled by the
Metropolis algorithm. MC autocorrelation times can be significantly
reduced by alternating Metropolis moves with microcanonical
over-relaxed ones~\cite{CCFTV2003,BrownW1987}. However, the real
improvement in simulation efficiency arises from the fact that the
move amplitudes can be independently chosen and dynamically adjusted
{\em for each} Fourier component, thus correctly sampling also
strongly fluctuating paths~\cite{CCFTV2002,WernerT2004}. Eventually,
the finite-$P$ overall action $S+S_{_{\rm{I}}}$ reads
\begin{equation}
 \sum_{\bm{ij}}\sum_{k=1}^M T_{\bm{ij}k}
 \Big(\varphi_{\bm{i}k}^{^{\rm{(R)}}}\varphi_{\bm{j}k}^{^{\rm{(R)}}}
 +\varphi_{\bm{i}k}^{^{\rm{(I)}}}\varphi_{\bm{j}k}^{^{\rm{(I)}}}\Big)
 - \frac1{2Pt}\sum_{\bm{id}}\sum_{\ell=1}^M\cos\varphi_{\bm{id}\ell}~,
\label{e.Sk}
\end{equation}
with the `kinetic' matrix $T_{\bm{ij}k}{=}\frac{P^2t}{g^2}\,
\sin^2\!{\frac{\pi{k}}P}\Gamma_{\bm{ij}}^{(\eta)}
{+}\gamma\,k~\Gamma_{\bm{ij}}^{(0)}~.$ The last term in
Eq.~(\ref{e.Sk}) containing the Josephson interaction is understood
to be expressed using the expansion~(\ref{e.Fphi}): note that it does
not burden the simulation as the $\{\varphi_{\bm{i}\ell}\}$ are
stored and just updated in the component to be moved, at variance
with the integral appearing in standard Fourier PIMC
algorithms~\cite{DollCF1985}. We performed an extensive set of
simulations involving $L\,{\times}\,L$ square lattices with linear
sizes up to $L\,{=}\,48$ and Trotter numbers up to $P\,{=}\,201$,
setting $\eta\,{=}\,0.01$.

\begin{figure}[t]
\includegraphics[width=85mm]{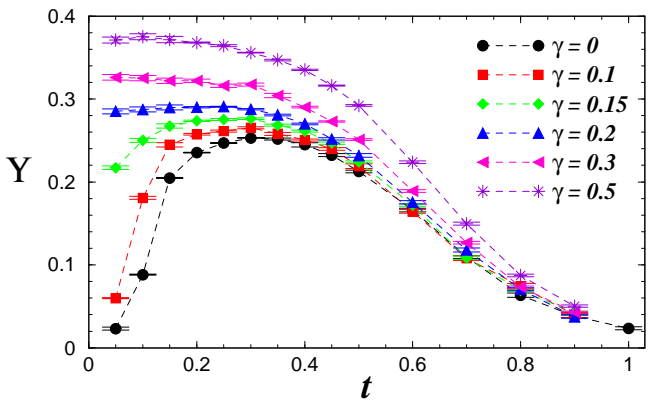}
\caption{
Temperature behavior of the helicity modulus $\Upsilon(t)$, on the
$8\,{\times}\,8$ lattice and $P\,{\to}\,\infty$, for $g\,{=}\,3.4$
and different dissipation strengths, showing that rising $\gamma$ the
reentrant behavior disappears and $\Upsilon(t)$ becomes monotonic for
$\gamma\gtrsim{0.2}$.
\label{f.2_Yt_gamma}}
\end{figure}

Our main results concern the phase diagram in the $(t,g)$ plane for
different values of the dissipation strength $\gamma$, and are
summarized in Fig.~\ref{f.1_Tg_gamma}; the data points were obtained
for $P\,{=}\,101$  as in Ref.~\onlinecite{CCFTV2003} by fitting the
finite-size scaling relation for the helicity modulus (or stiffness)
per island $\Upsilon=(L^2\,E_{_{\rm{J}}})^{-1}\,
\big[{\partial^2{F}}/{\partial{q^2}}\big]_{q=0}$~,
defined as the response of the free energy $F(q)$ under twisting the
boundary conditions as
$\varphi_{\bm{i}}\to\varphi_{\bm{i}}+\bm{q}{\cdot}\bm{i}$.

As shown in Fig.~\ref{f.1_Tg_gamma}, in addition to generally
stabilize the S phase, dissipation hinders the mechanism causing the
reentrance. However, the reentrance persists for small values of
$\gamma$ and a finite value $\gamma\,{\gtrsim}\,0.2$ is necessary to
restore a monotonic critical line. The persistence of a reentrant
normal phase for $\gamma\,{=}\,0.15$ is illustrated in the inset of
Fig.~\ref{f.1_Tg_gamma}, where the thermodynamic helicity modulus, as
obtained through a systematic finite-size scaling analysis, is
plotted as a function of the quantum coupling: at $g\,{=}\,3.5$ it
appears that $\Upsilon$ keeps a finite value for $t\,{=}\,0.20$,
while it vanishes for $t\,{=}\,0.05$, thus signaling that the
established phase coherence disappears again at low temperature.

It is interesting to see how the temperature behavior of
$\Upsilon(t)$ changes with $\gamma$ in the region of the reentrance.
In Fig.~\ref{f.2_Yt_gamma} several finite-lattice data for $g=3.4$
show that rising $\gamma$ removes the reentrance to disorder at low
temperature, as if $g$ were decreased~\cite{CCFTV2003}. From these
data one can again roughly estimate that the threshold where the
reentrance disappears is at $\gamma\,{\simeq}\,0.2$.

Further increasing $\gamma$ the critical line in
Fig.~\ref{f.1_Tg_gamma} progressively changes its curvature and a
flex appears for $\gamma\,{\gtrsim}\,0.4$, signaling the incipient
stabilization of a low-temperature S phase for any value of the
quantum coupling, as we argue below. In order to address this
important issue, we study the dependence on $\gamma$ of the helicity
modulus and related quantities at higher values of $g$, namely
$g\,{=}\,5$ and $g\,{=}\,10$, and at a fixed low temperature
$t\,{=}\,0.05$.
\begin{figure}[t]
\includegraphics[width=85mm]{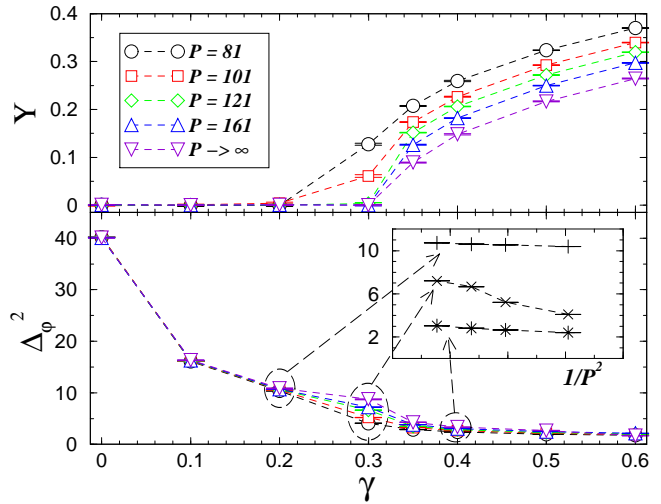}
\caption{Finite-size data ($L\,{=}\,24$) for $\Upsilon$ (top panel) and
$\Delta_\varphi^2$ (bottom panel) vs $\gamma$, at $t\,{=}\,0.05$ and
$g\,{=}\,5$, for different values of $P$. Inset: $1/P^2$
extrapolation of $\Delta_\varphi^2$.
\label{f.3_Yalpha_gamma}}
\end{figure}
In Fig.~\ref{f.3_Yalpha_gamma} we report the helicity modulus and the
`pure-quantum' spread of the phase difference between neighboring
islands~\cite{HerreroZ2002,CCFTV2003}, namely
$\Delta^2_\varphi=\big\langle\,
[\varphi_{\bm{id}}(u){-}\bar{\varphi}_{\bm{id}}]^2\,\big\rangle$, for
$g\,{=}\,5$ and different Trotter numbers including the extrapolation
to $P\,{\to}\,\infty$. For increasing dissipation, $\Upsilon$ remains
zero in an interval and then, around a crossover value
$\gamma\,{\simeq}\,0.3$, it abruptly starts to increase; for slightly
larger $\gamma$ the critical curve $t{_{_{\rm{C}}}}(g,\gamma)$ is
hence expected to cross the point $(g\,{=}\,5,\,t\,{=}\,0.05)$. The
crossover is clearly connected with the quenching of the pure-quantum
phase fluctuations (lower panel of Fig.~\ref{f.3_Yalpha_gamma}). Just
at the crossover, the PIMC data for $\Delta_\varphi^2$ display the
phenomenon of a markedly weaker convergence with $P$ (see figure
inset), signaling that fluctuations of high Matsubara modes become
significant. We suggest that this reflects the proximity to a phase
transition mainly driven by quantum fluctuations, which are in turn
modulated by the dissipation.

\begin{figure}[t]
\includegraphics[width=85mm]{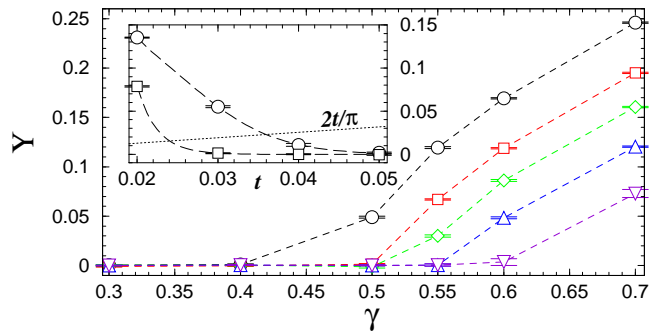}
\caption{
Finite-size data ($L\,{=}\,24$) for $\Upsilon$ vs $\gamma$, at
$t\,{=}\,0.05$ and $g\,{=}\,10$, for different values of $P$ as in
Fig.~\ref{f.3_Yalpha_gamma}. Inset: $\Upsilon(t)$ for
$\gamma\,{=}\,0.55$ (circles) and $\gamma\,{=}\,0.50$ (squares) for
$L\,{=}\,12$ and $P\,{=}\,201$. The straight dotted line in the inset
marks the universal-jump value $2t/\pi$.
\label{f.4_Y_gamma}}
\end{figure}

Since at $g\,{=}\,5$ the system displays a BKT transition for
$\gamma\,{\gtrsim}\,0.3$, a much larger coupling is required to check
whether $\gamma\geq\gamma_{_{\rm{C}}}\,{=}\,1/2$ in fact ensures
low-$t$ ordering. We therefore performed simulations for quantum
coupling as large as $g\,{=}\,10$, as reported in
Fig.~\ref{f.4_Y_gamma}. Again a crossover of $\Upsilon$, signaling
the proximity of the BKT critical line, shows up at
$\gamma\,{\simeq}\,0.6$: as $t$ is still finite, this value is only
an upper bound for $\gamma_{_{\rm{C}}}$. Indeed, as shown in the
inset, the transition  occurs at low $t$ also for $\gamma\,{=}\,0.5$.
Our results are therefore consistent with
$\gamma_{_{\rm{C}}}\,{=}\,1/2$, in agreement with early
predictions~\cite{ChakravartyIKL1986,Fisher1987,ChakravartyIKZ1988}.
In addition, the reentrance displayed at $\gamma\,{=}\,0$ disappears
well before $\gamma{_{_{\rm{C}}}}\,{=}\,1/2$, so there is no evident
connection between the two phenomena.

\begin{figure}[t]
\includegraphics[width=85mm]{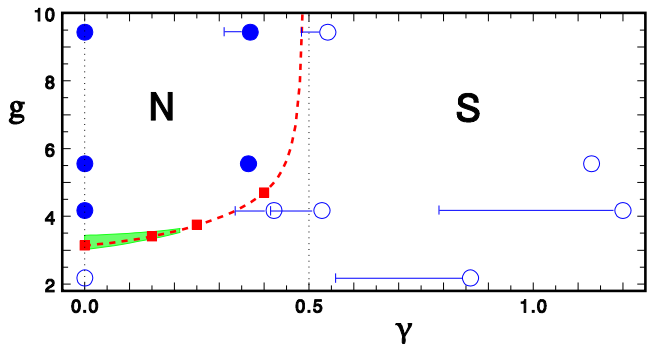}
\caption{
Zero-$t$ phase diagram. Squares: estimates of $g_{_{\rm{C}}}$ from
the data reported in Fig.~\ref{f.1_Tg_gamma}. The dashed line
extrapolates them to the expected behavior of
$g_{_{\rm{C}}}(\gamma)$. Full and open circles: N and S phase,
respectively, as observed experimentally~\cite{TakahideYKOK2000}. The
reentrant behavior occurs in the shadowed region.
\label{f.5_pd_t0}}
\end{figure}

The quantum phase transition occurs at the value
$g\,{=}\,g_{_{\rm{C}}}(\gamma)$ where the critical temperature
$t_{_{\rm{C}}}(g,\gamma)$ vanishes: assuming that
$\gamma_{_{\rm{C}}}\,{=}\,1/2$ and using the data reported in
Fig.~\ref{f.1_Tg_gamma} to estimate $g_{_{\rm{C}}}$ for some values
of $\gamma$, we can draw the line of quantum critical points
$g_{_{\rm{C}}}(\gamma)$ that separates the S and N phase in the
$(\gamma,g)$ plane. Fig.~\ref{f.5_pd_t0} evidences that the resulting
zero-$t$ phase diagram is in remarkable agreement with the
experimental findings by Takahide {\it et
al.}~\cite{TakahideYKOK2000}.

In conclusion, we have obtained the quantitative phase diagram of a
resistively shunted 2D Josephson junction array, for several
dissipation strengths $\gamma$. Our results indicate that the
reentrant low-temperature normal phase, recently evidenced in the
limit of infinite shunt resistance
($\gamma\,{\to}\,0$)~\cite{ZantEGM1996,CCFTV2003}, persists in a
small but {\em finite} range of values of $\gamma\,{\lesssim}\,0.2$
(i.e., $R_{_{\rm{S}}}\,{\gtrsim}\,32$\,k$\Omega$) and
$3.2\,{\lesssim}\,g\,{\lesssim}\,3.5$. This explains why the
reentrance was not detected in the shunted JJA samples of
Ref.~\onlinecite{TakahideYKOK2000}, displaying resistances
$R_{_{\rm{S}}}\,{\lesssim}\,18$\,k$\Omega$ and quantum coupling
values ($g\,{=}\,2.2$, $4.2$, $5.5$, $9.4$) outside the above range.
For $\gamma\,{>}\,\gamma_{_{\rm{C}}}\,{=}\,1/2$ we observe the SN
transition at finite $t$ for very large coupling, so that our data
validate the prediction that above $\gamma_{_{\rm{C}}}$ the
superconducting phase is always stabilized. Our extrapolated
zero-temperature phase diagram explains the available experimental
observations.

We thank J.~V. Jos\'e and Y.~Takahide for fruitful correspondence.
V.~T. acknowledges the hospitality at the `Abdus Salam'
ICTP--Trieste. This work has been supported by the COFIN2002-MIUR
fund.

\end{document}